\documentstyle[epsfig,12pt]{article}
\begin{document}
\begin{quote}
\raggedleft
%hep-ph/yymmxxx \\
PM/97-10 \\
May 15, 1997
\end{quote}

\vspace*{3cm}
\begin{center}
{\bf ANALYTIC CONSTRAINTS\\
\vspace{-0.3cm}
 FROM EW SYMMETRY BREAKING IN THE MSSM}\footnote{supported
in part by EC contract CHRX-CT94-0579}
\end{center}
\setlength\baselineskip{12pt}
\begin{center}
%\vspace{-0.3cm}
\end{center}
\begin{center}
Gilbert MOULTAKA
\end{center}
\vspace{-0.5cm}
\begin{center}
{\it Physique Math\'ematique et Th\'eorique\\
UPRES-CNRS A 5032,\\
Universit\'e Montpellier II, F34095 Montpellier Cedex 5, France}
\end{center}
%\vspace{3.5cm}
\begin{center}
{\sl (Talk given at XXXIInd Rencontres de Moriond- 15-22 March 1997,\newline
ELECTROWEAK INTERACTIONS AND UNIFIED THEORIES)}
\end{center}
\vspace{3cm}
\begin{abstract}
\setlength\baselineskip{12pt}
We report on how a straightforward (albeit technically involved) analytic
study of the 1-loop effective potential in the Minimal Supersymmetric
Standard Model (MSSM), modifies the usual electroweak symmetry breaking
(EWSB) conditions involving $\tan \beta$ and the other free parameters
of the model. The study implies new constraints which ( in contrast
with the existing ones like $1 \leq \tan \beta \leq m_t/m_b$ ) 
are fully  model-independent and exclude more restrictively a region around
$\tan \beta \sim 1$. Further results of this study will be only touched upon 
here.
\end{abstract}
\vspace{3cm}
\begin{center}
{\sl (to appear in the proceedings)}
\end{center}

\setlength\baselineskip{15pt}

\newpage
Determining the structure of the true vacuum is a central task in particle 
physics. It encodes the mass spectrum, the way symmetries are broken and 
elementary fields interact. Supersymmetric theories give specific properties
to the vacuum structure, some of which survive or remain approximately true
after the soft breaking of supersymmetry. An attractive feature of this 
symmetry breaking is that it can trigger (together with a heavy top quark), 
the spontaneous breaking of the electroweak symmetry \cite{EWB}. However, 
such a phenomenon should be accompanied by some extra constraints to ensure 
that neither charge/color breaking nor unbounded from below potential occur 
\cite{casas}. In the absence of any underlying mechanism that dictates those
constraints, one usually resorts to a phenomenological determination using the
approximation at hand of the true effective potential of the theory. It is
then useful to keep in mind some essential features: \newline 
{\sl i)} The true effective potential $V_{eff}$ should be renormalization scale invariant. {\sl ii)} The only gauge parameter independent and physically 
relevant values of $V_{eff}$ are those at stationary points. {\sl iii)} 
the renormalization group (RG) improved effective potential 
$\overline{V}_{eff}$
can miss large logarithms when there are more than one (mass) scale in the 
theory, as is the case in the MSSM. If not properly treated, the scale at which
leading logs are actually resummed becomes fuzzy and with it the approximate
scale invariance of $\overline{V}_{eff}$. {\sl iv)} A local minimum is 
characterized by both $1^{st}$ {\it and} $2^{nd}$ order derivatives 
with respect to {\it all} 
scalar fields in the theory. {\sl v)} Non logarithmic contributions to 
$V_{eff}$ at a given loop order can alter the qualitative picture
of the ``tree-level improved'' effective potential.\\
In this short presentation we concentrate on the interplay between {\sl iv)} 
and {\sl v)}, which leads, as we will see, to a modification of the usual 
discussion of EWSB
conditions in the MSSM.

The 1-loop effective potential has the well-known form \cite{CW}
\begin{equation}
V= V_{tree} + \frac{\hbar}{64 \pi^2} Str[ M^4 (Log \frac{M^2}{\mu_R^2} -
3/2) ] \label{EP}
\end{equation}
in the $\overline{MS}$ scheme. Here $\mu_R$ denotes the renormalization scale, 
$V_{tree}$ the tree-level MSSM potential
\cite{susy}, $M^2$ the field dependent squared mass matrix of the scalar 
fields,
and $Str[...] \equiv \sum_{spin} (-1)^{2 s} (2 s + 1) (...)_s $, summing 
over gauge boson, fermion and scalar contributions. 
In the MSSM the explicit expression of
$Str[ M^4 (Log \frac{M^2}{\mu_R^2} -3/2)]$ is overwhelmingly 
lengthy. We have determined its exact structure analytically,  relying on
symbolic computation, thus allowing the exploration of virtually any direction
in field space \cite{nous}. Hereafter we will be  interested in the 
$(H_1, H_2)$ manifold, 
{\it i.e.} all scalar fields but the two Higgs doublets are put to zero; 
$ H_1\equiv ( H_1^0, H_1^{-}), H_2 \equiv ( H_2^{+}, H_2^0)$ consist of 8 real
valued fields. Furthermore we limit the study here to $Str [ M^4]$, that is we
assume that 
logarithms in eq.(\ref{EP}) are reabsorbed in the running of all the parameters
in $\overline{V}_{tree}(\mu_R^2$) \setlength\baselineskip{12pt}
\footnote{
bypassing difficulties related to points {\sl i)} and  {\sl iii)} above, 
as is often
implicitly done in the literature! see however \cite{multiscale}.}.
\setlength\baselineskip{15pt} 
From this point of view $(-3/2)\hbar/64 \pi^2 Str[M^4]$ is 
the genuine residual 1-loop effect. The effective potential takes then
the form
\begin{equation}
 V= \overline{V}_{tree}(\mu_R^2) + \frac{\hbar}{64 \pi^2} (-3/2) Str M^4 
\label{Poteff}
\end{equation}
Now comes the main point: the usual EWSB conditions \cite{EWB},
\begin{eqnarray}
\frac{1}{2} M_Z^2 = \frac{\overline{m}_1^2 - \overline{m}_2^2 \tan^2 \beta}
{\tan^2 \beta -1 }  &,&
\sin 2\beta= \frac{2 \overline{m}_3^2}{\overline{m}_1^2 + \overline{m}_2^2}
\label{EWSBcond} \\
\nonumber
\end{eqnarray}
being merely the expressions of vanishing {\it first order} derivatives in the
neutral directions \newline $<H_1>=(v_1, 0), <H_2>=(0, v_2) $, 
do not as such guarantee 
that we have indeed a (local) minimum of the effective potential at the 
electroweak scale, see point {\sl iv)} above. However,  provided that the 
$\overline{m}_i's$ are $v_1$ and $v_2$ independent, one can show that
eq.(\ref{EWSBcond}), supplemented with the boundedness-from-below condition
$\overline{m}_1^2 + \overline{m}_2^2 - 2 |\overline{m}_3^2| \geq 0$ does 
indeed imply that the 
8 eigenvalues of the {\sl second order} derivatives matrix are positive.
It is noteworthy that this connection between $1^{st}$ and $2^{nd}$ order
derivatives holds at tree-level, due to the special form of $V_{tree}$ as 
dictated by softly broken supersymmetry as well as to the fact that the
$\overline{m}_i$'s are all trivially $v_1, v_2$ independent at this level.
It also continues to hold for the same reasons in the tree-level improved
approximation ($\overline{V}_{tree}(\mu_R^2)$) where the running 
$\overline{m}_i(\mu_R^2)$ are $v_1, v_2$ independent as well ${}^2$.
[This is why eqs.(\ref{EWSBcond}) are often referred to in the literature
as the {\sl minimization conditions}, although they are not intrinsically so.] 
Nonetheless, there is no reason to believe that such a connection will
hold for the full theory. Actually as we will see,
it will already break down when non logarithmic 1-loop corrections are 
included, eq.(\ref{Poteff}). 
Indeed in this case, even though eqs.(\ref{EWSBcond}) retain their generic 
form,
through a redefinition of the $\overline{m}_i$'s, there is now an explicit 
dependence on $v_1$ and $v_2$ in the $\overline{m}_i's$. Injecting the full 
$H_1, H_2$ contribution of $Str M^4$ and extracting this explicit dependence, 
one finds the conditions for vanishing $1^{st}$ order derivatives: 
\vspace{-0.3cm}
\begin{eqnarray}
X_{m_1}^2 v_1 - X_{m_3}^2 v_2 + X v_1 ( v_1^2 - v_2^2) + \tilde{\alpha} v_1^3 = 0  &,&
X_{m_2}^2 v_2 - X_{m_3}^2 v_1 + X v_2 ( v_1^2 - v_2^2) - \tilde{\alpha} v_2^3 = 0  \nonumber \\
\label{EWSBnew}
\end{eqnarray}
Here $X_{m_i}= m_i + \hbar \delta m_i$ and $X=g^2/8 + \hbar \delta g$, 
where $\delta m_i$ (resp. $\delta g$)  have been determined analytically 
and depend 
on all mass parameters of the MSSM and Yukawa couplings 
(resp. gauge couplings).
 $\tilde{\alpha} \equiv -9\hbar/(256\pi^2) g^2(Y_t^2 - Y_b^2)$ is a purely 
residual one-loop contribution. [ $g^2 \equiv g_1^2 + g_2^2$, $g_1$ 
(resp. $g_2$) being the $U(1)$ (resp. $SU(2)_L$) gauge couplings, 
and we show in 
$\tilde{\alpha}$ just the top/bottom Yukawa dependence]. 
It is important to note
at this level that a non zero $\tilde{\alpha}$ in eq.(\ref{EWSBnew}) modifies
drastically the $\tan \beta$ dependence in the EWSB conditions. While these 
conditions are quadratic in $\tan \beta$  when $\tilde{\alpha} = 0$, 
(leading to eqs.(\ref{EWSBcond}) with $v_1, v_2$ 
free $\overline{m}_i$'s ), 
they become quartic in $\tan \beta$ (and linear in $u \equiv v_1 v_2$)
when  $\tilde{\alpha} \neq 0$. More importantly now, one can show analytically 
that eqs.(\ref{EWSBnew}) 
together with the 1-loop boundedness-from-below condition 
$X_{m_1}^2 + X_{m_2}^2 - 2 |X_{m_3}^2| \geq 0$ no
more suffice to characterize a local minimum and one has to resort explicitly
to the positiveness conditions of all the physical Higgs squared 
masses\footnote{
Three of the eight eigenvalues are of course still vanishing and correspond
to the goldstone components.}. However, working out the details \cite{nous},
it turns out that only two independent new conditions are required and read 
when coupled with eqs.(\ref{EWSBnew})
(we give here for conciseness an approximate form of the first condition) :
\vspace{-0.1cm}
\begin{equation}
\mbox{ {\sl a)}\,}  \tan\beta \leq t_{-} \mbox{\,or\,} \tan\beta \geq t_{+} 
\end{equation}
\begin{equation}
\mbox{\,where\,\,} t_{\pm} \simeq 1 \pm \frac{4}{g} 
\sqrt{\frac{|\tilde{\alpha}| 
u} {X_{m_3}^2}} \sqrt{|\tilde{\alpha}|} + O(\tilde{\alpha})
\end{equation}
\vspace{-0.3cm}
\begin{equation}
\mbox{ {\sl b)} \,} \frac{X_{m_3}^2}{u} \geq 0 \Leftrightarrow 
\tan \beta\geq T_{+} \,(\mbox{\, if \,} \tan \beta > 1) \mbox{\, or  \,} 
\tan \beta\geq T_{-} \,(\mbox{\, if \,} \tan \beta < 1) 
\end{equation}
\begin{equation}
\mbox{\,where\,\,}
T_{\pm} = \frac{ X_{m_1}^2 + X_{m_2}^2 \pm \sqrt{(X_{m_1}^2 + X_{m_2}^2)^2 -
4  X_{m_3}^4 }}{2 X_{m_3}^2} \nonumber
\end{equation}

A couple of remarks are in order. Eqs.(\ref{EWSBnew})
and conditions {\sl a)} and {\sl b)} above, are necessary and sufficient 
conditions to 1-loop order for a local minimum to occur in the neutral
$<H_1>=(v_1, 0), <H_2>=(0, v_2) $ direction.
This is due to that on one hand a charge breaking vacuum does not occur
in the $H_1, H_2$ sector (even to 1-loop order \cite{nous}), 
contrary to the non supersymmetric two-higgs-doublet case \cite{sher}, 
and on the other hand, eqs.(4-8)
imply by themselves that zero field configurations are energetically unstable,
thus leading to spontaneous symmetry breaking. 
The boundary values $t_{\pm}$ in eq. (6) depend implicitly on $\tan \beta$
through $u$, the latter being determined by solving 
eqs.(\ref{EWSBnew}), where $\tilde{\alpha} u$ is formally $O(1)$ in 
$\tilde{\alpha}$. Furthermore, one has generically $T_{+} > t_{+}$ and
$T_{-} < t_{-}$. 
Conditions {\sl a)} and {\sl b)} exclude altogether a {\sl calculable}
region around $\tan \beta \sim 1$ in the otherwise unconstrained general MSSM. 
This should be contrasted with the usually quoted model-dependent constraint 
$1 \leq \tan\beta (\leq
m_t/m_b) $ which requires an assumption of universality at the unification
scale \cite{giudice}\footnote{ the other argument in favour of 
$1 \leq \tan\beta (\leq m_t/m_b) $, sometimes quoted in connection with
perturbativity, is less quantitative and suffers from a less physical 
setting .}.
It is also reassuring that in the limit $\hbar \to 0$, 
the values of $T_{\pm}$ degenerate into the $\tan \beta$ solutions for 
eqs.(\ref{EWSBcond}) while $t_{\pm} \to 1$, so that {\sl a)} and {\sl b)} 
are not constraints any more in this limit, as it should in the tree-level 
approximation.\\[-5mm]

Although valid in a model-independent context,
it is obvious that conditions {\sl a)} and {\sl b)} can also be included 
consistently in a SUGRA-GUT motivated analysis. For instance in the 
1-loop order approximation,
one runs just the tree-level parts of the $X_{m_i}$'s and $X$, from 
say a unification scale
downwards, 
while their 1-loop residual parts and
$\tilde{\alpha}$ do not run. Eqs.(\ref{EWSBnew}) are then solved for 
$\tan \beta$ and $u$ and the solutions not fulfilling eqs.(5-8) eliminated.
It should be clear that this is done {\sl prior} to any other physical 
requirement like reproducing the correct $M_Z$ at the relevant scale, 
constraints on $M_{top}$, proton life-time, etc..., 
and as such could have a feedback on these requirements.\\
Finally, even if the requirement of positive squared physical Higgs masses
can be accounted for in a numerical analysis, we have established here
analytically its theoretical connection with a model-independent constraint
on the allowed values of $\tan \beta$.\\

This work was done in collaboration with Christophe Le Mou\"el.
I would like to thank Gordon Kane, Dmitri Kazakov and Nikolai Krasnikov
for suggestions and useful discussions during the conference. 
\vspace{-1.cm} 
\setlength\baselineskip{10pt}

\end{document}